\documentclass[osajnl,twocolumn,showpacs,superscriptaddress,10pt]{revtex4-1}

\usepackage{amsmath,amssymb,graphicx}

\begin{document}

\title{Simple and robust speckle detection method for fire and heat detection in harsh environments}

\author{Charles N. Christensen}
\affiliation{Department of Chemical Engineering and Biotechnology, University of Cambridge, Cambridge, UK}
\affiliation{Danish Fundamental Metrology, Kogle Alle 5, DK-2970 H{\o}rsholm, Denmark}

\author{Yevgen Zainchkovskyy}
\affiliation{Danish Fundamental Metrology, Kogle Alle 5, DK-2970 H{\o}rsholm, Denmark}

\author{Salvador Barrera-Figueroa}
\affiliation{Danish Fundamental Metrology, Kogle Alle 5, DK-2970 H{\o}rsholm, Denmark}

\author{Antoni Torras-Rosell}
\affiliation{DPA Microphones A/S, Gydevang 42-44, 3450 Lillerod Denmark}

\author{Giorgio Marinelli}
\affiliation{Dansk Brand- og Sikringsteknisk Institut, Jernholmen 12, 2650 Hvidovre, Denmark}

\author{Kim Sommerlund-Thorsen}
\affiliation{Dansk Brand- og Sikringsteknisk Institut, Jernholmen 12, 2650 Hvidovre, Denmark}

\author{Jan Kleven}
\affiliation{Elotec AS, S{\o}ndre Industrivegen 1, NO-7340 OPPDAL, Norway}

\author{Kristian Kleven}
\affiliation{Elotec AS, S{\o}ndre Industrivegen 1, NO-7340 OPPDAL, Norway}

\author{Erlend Voll}
\affiliation{Elotec AS, S{\o}ndre Industrivegen 1, NO-7340 OPPDAL, Norway}

\author{Jan C. Petersen}
\affiliation{Danish Fundamental Metrology, Kogle Alle 5, DK-2970 H{\o}rsholm, Denmark}

\author{Mikael Lassen}
\affiliation{Danish Fundamental Metrology, Kogle Alle 5, DK-2970 H{\o}rsholm, Denmark}
\affiliation{ml@dfm.dk}

\begin{abstract}
Standard laser based fire detection systems are often based on measuring variation of optical signal amplitude. However, mechanical noise interference and loss from dust and steam can obscure the detection signal, resulting in faulty results or inability to detect a potential fire. The presented fire detection technology will allow the detection of fire in harsh and dusty areas, which are prone to fires, where current systems show limited performance or are unable to operate. It is not the amount of light nor its wavelength that is used for detecting fire, but how the refractive index randomly fluctuates due to the heat convection from the fire. In practical terms this means that light obstruction from ambient dust particles will not be a problem as long as a small fraction of the light is detected and that fires without visible flames can still be detected. The standalone laser system consists of a Linux-based Red Pitaya system, a cheap 650 nm laser diode, and a PIN photo-detector. Laser light propagates through the monitored area and reflects off a retroreflector generating a speckle pattern. Every 3 seconds time traces and frequency noise spectra are measured and 8 descriptors are deduced to identify a potential fire. Both laboratory and factory acceptance tests have been performed with success. 
\end{abstract}

\maketitle

\section{Introduction}

State-of-the-art sensors for fire detection are fairly slow, as they for the most cases rely on the detection of gases and particles that reach a certain threshold concentration \cite{Li2001, Knox2011}, which occurs only long after the initial fire has started \cite{Evans1986,Lee1974}. Additionally, standard smoke detectors produce frequent false alarms, which can result in unnecessary shutdowns, loss of operational continuity, being costly, cause unscheduled system replacements and high service costs \cite{Chagger2014}. False alarms may be set off by particles not related to fire, like dust or pollution particles, which are common in industrial plants. This conditions users to ignore real fire alarm systems causing, in worst-case scenarios, considerable losses, including lives. Furthermore, conventional smoke detectors have to be mounted just below ceilings. The higher the room is the bigger the fire has to be to trigger the detector. In large atrium and other buildings with a high ceiling, this means that the alarm notifies when the smoke has already spread to the whole room. By dealing with this threat at an early stage, the damage and subsequent losses can be significantly reduced, and it is therefore very important to detect and react as quickly as possible. Another drawback for smoke detectors is in identifying fires from fuels which only generate a small amount of smoke.

Optical and laser based systems have shown great potential to solve these drawbacks. Optical and Laser based fire detection systems have been known for many years \cite{Lawson1966}. The common way to recognize a fire is by direct monitoring the radiation from flames, heat, gasses or particulates released during the burning process. Generally the systems comprises a laser source or any illumination source, which emits a light beam that crosses the region to be monitored and a detector that receives the light beam after having passed the monitored region. The received light beam is then analyzed to determine if a fire has occurred. Most of these sensors still rely on the fact that light attenuation may limit the performance and give false alarms. Many laser based fire detection systems are based on measuring or imaging changes in light intensity (signal amplitude), either due to heat flow, smoke and dust or using laser spectroscopic of fire relevant molecules or a combination of all the techniques \cite{Lancaster1999,Zeller2010,Milke1995,Umar2017,Chen2007}. Since fire is an exothermic reaction, all fires will release heat, which results in random changes of the refractive index of the air \cite{Atkinson1992}. Fire generates random fluctuations in the refractive index due to convective flow of heat along the beam´s propagation path \cite{Hollman1976}, which creates random fluctuations in the refractive index, resulting in beam distortion, phase changes and beam displacement and tilt. The temporal and spatial variation of the laser beam can therefore be measured as intensity variations, phase changes, beam wander, or beam spread changes \cite{Lee1974,Isterling2012,Umar2017,Delaubert2006}. Different reports have been made on heat/fire detectors that sense the movement of hot gases due to the fire causing refraction of the beam either toward or away from the optical detector. If the beam is normally incident on the detector (i.e. in the absence of fire), movement of the beam away from detector, due to refraction, is sensed.

We present a complete new solution for the detection of fire, which is an efficient, fast and relatively low cost method for use in larger harsh areas, such as industrial sites and/or large constructions. The method is based on speckle pattern intensity modulation and does not rely on the absolute intensity of the light beam, making the sensor immune to general attenuation due to dust and smoke. Speckle patterns are an intensity pattern produced by the mutual interference of monochromatic light incident on a rough surface \cite{Goodman1976}.  In general speckle pattern that changes in time, due to changes in the illumination is known as dynamic speckle, and are for example used for optical flow sensors, optical computer mouses  and  for biological materials, known as biospeckle \cite{Almoro2009,Nassif2014}. We present a new application of dynamical speckle pattern measurements for the detection of fire. The generation of speckle patterns is dependent on the beam wavelength, size and beam incident angle and the roughness of the reflecting surface. However, the angle and distance to the observer also matters for the amount of detected back scattered light. When a fire is occurring the heat convection will generate a randomly fluctuating changes in the refractive index, which will make the laser beam jitter (perform random chaotic walk) and/or defocus the beam on the reflective rough surface, thus generating an intensity modulation of the optical power as a function of the random heat convection. The retro-reflected laser light (speckle pattern) is then simply collected by an optical detector and the noise power spectrum in a broad frequency range is calculated. Broad frequency noise (from 10 Hz -5 kHz) is an indication of fire, while on the other hand narrow linewidth noise sources are indication of mechanical resonances. Which requires reference noise traces of the background noise without any fire. The system is able to identify very small temperature fluctuation which rises at a very early stage of a fire, providing an early and reliable detection of fire. Measurements have been performed for ranges of up 100 meters in both a controlled laboratory environment and at a factory (Energnist I/S waste plant in Kolding, Denmark). The aim of the research is to develop a fire detection system that is able to identify very small temperature rises at a very early stage of a fire, thus providing an early and reliable detection of fire, which will decrease the risk of false alarms. The final system will target applications with challenging environments such as waste management facilities, power supply plants, piggeries, cow stables, food processing factories and other dusty locations, where traditional heat point detectors struggle and cannot survive more than a few years. Furthermore, the final advanced fire detection system is also conceived to be installed in places where line smoke detectors – like beam detectors – cannot be used today due to e.g. the presence of particles from combustion, like tunnels, engines rooms, cargo halls, metro stations, and ships. To our knowledge the sensor is the first of its kind and a completely new solution for the fire detection market.

\section{Random fluctuation of the refractive index and speckle patterns}

\begin{figure}[htbp]
\centering
\fbox{\includegraphics[width=\linewidth]{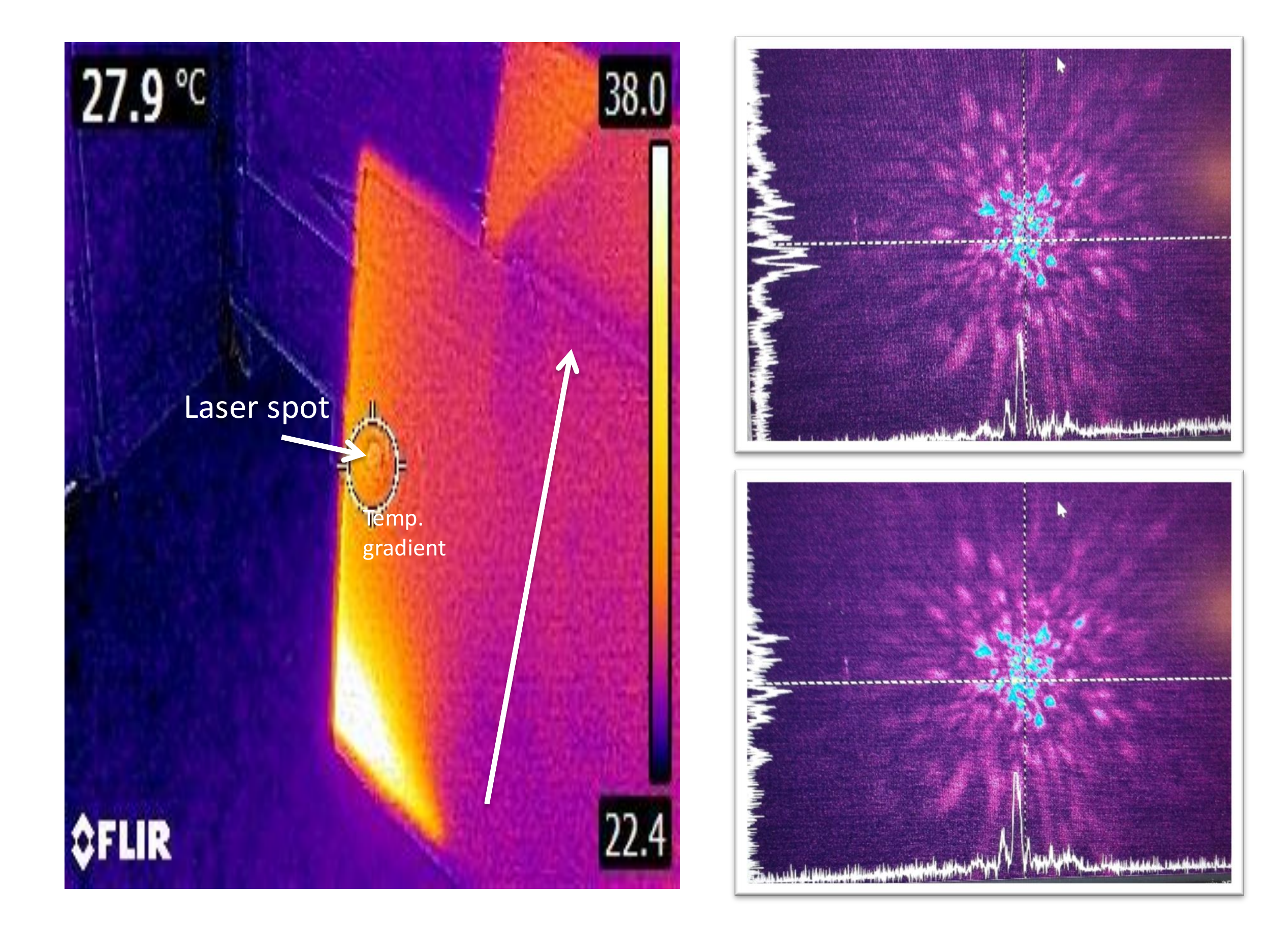}}
\caption{Left image:FLIR images of the heat flow on a 1 cm thin cardboard plate. For visualization of the heat convection. Temperature of the measurement point (marked with the circle). The sensor was placed 10 meter away. The temperature gradient is clearly seen and is indicated with the arrow. The measurement point has a temp of approximately 28$^\circ$C, while 25 cm below the laser spot the temp is approximately 38$^\circ$C. The heat flow is generated with a heat gun situated 1 m below the surface. The heat gun has a temperature of approximately 260$^\circ$C. Right images: Measured speckle patterns with a beam profiler with 5 second separation.}
\label{fig2}
\end{figure}

The fire detection technique relies on the measurement of intensity modulation of speckle patterns due to heat convection. Since only a small part of the back reflected speckle patterns are measured the change will give the needed intensity modulation. Note that having a highly sensitive detector and measuring a single speckle interference spot will in principle give a interferometric visibility of 100 $\%$, thus having the highest SNR from an optical point of view, unfortunately the electronical noise of the photo-detector will be dominating in most cases. Figure \ref{fig2} shows an image of a typical measurement situation, where a heat gun is used to simulate a fire. The laser measurement point has a temperature of approximately 28$^\circ$C, while 25 cm below the laser spot the temperature is approximately 38$^\circ$C. The heat gun has a temperature of approximately 260$^\circ $C. The heat convection can be clearly seen and in the figure shown as arrow for visualization. Figure \ref{fig2} also shows the measured speckle patterns measured with 5 seconds separation. When a fire is occurring the heat convection will generate a randomly fluctuation of the refractive index, as also shown in Figure \ref{fig2}, which will make the laser beam jitter and perform random chaotic walk on the reflective rough surface \cite{Buck1967,Chiba1971}. This generates an intensity modulation of the speckle pattern as a function of the random heat convection. However, changes in the refractive index is implicit in both pressure $p$ and temperature $T$, which makes it difficult to decouple the effect from steam, vapor or pressure release from  industrial machines. Ambient temperature drifts are easy to decouple as they take place at a very slow pace, meaning that their energy is below the frequency range of interest (below 10 Hz). In order to separate these two contributions, the following first order Taylor expansion can be used \cite{Atkinson1992, Hill1980}:

\begin{equation}
n \cong n_0 + \left(\frac{\partial n}{\partial p}\right)_0 p' + \left(\frac{\partial n}{\partial T}\right)_0 T',
\label{eq:n_taylor_p_T}
\end{equation}
where $T'$ is the ambient temperature changes and $p' = p-p_0$ is the acoustic pressure, where $p_0$ is the static pressure. At $T_0=20 ^\circ$C, and for $p_0=101.325$~kPa (assuming a humidity percentage $H = 50$\% and a concentration of CO$_2$ of 0.05\%) the following values for $(\partial n/\partial p)_0$ and $(\partial n/\partial T')_0$ are obtained:

\begin{equation}
\left(\frac{\partial n}{\partial p}\right)_0 = 1.914\times 10^{-9}~Pa^{-1}\qquad\left(\frac{\partial n}{\partial T'}\right)_0 = - 9.567\times10^{-7}~^\circ C^{-1}
\label{eq:2}
\end{equation}
From the equation it can be seen that $(\partial n/\partial p)_0$ is about 500 times smaller than $(\partial n/\partial T)_0$. This means that conventional laser based interferometric techniques are considerably more sensitive to temperature changes than to pressure fluctuations. An immediate challenge for the sensing method is to determine whether a change in the frequency power noise spectrum is caused by a fire or the change is caused by a non-fire event. Non-fire events include influence of sunlight and the presence of ventilation and heating systems. In the beginning of a fire, the temperature changes are very small and it take minutes before a fire has envolved, there is therefore a tradeoff in system sensitivity and stability. In essence, the more sensitive the system is configured to be, the higher the risk of a false alarm. As a result a key task is to determine if a change in the frequency spectrum is caused by a non-fire event. The implementation and development of the demodulation algorithm has to make it possible to extract the early fire detection profile. This includes the signal processing modules as well as the software used to analyse the temperature changes inferred from the laser beam. The software needs to handle the time traces and frequency spectrum of the measured signal so that it can distinguish between fire and non-fire situations. Additional filters and algorithms need to be developed to deduct the baseline temperature profile to remove temperature changes based on slowly heating or cooling the room.”

\section{Experimental setup and results}

\begin{figure}[htbp]
\centering
\fbox{\includegraphics[width=\linewidth]{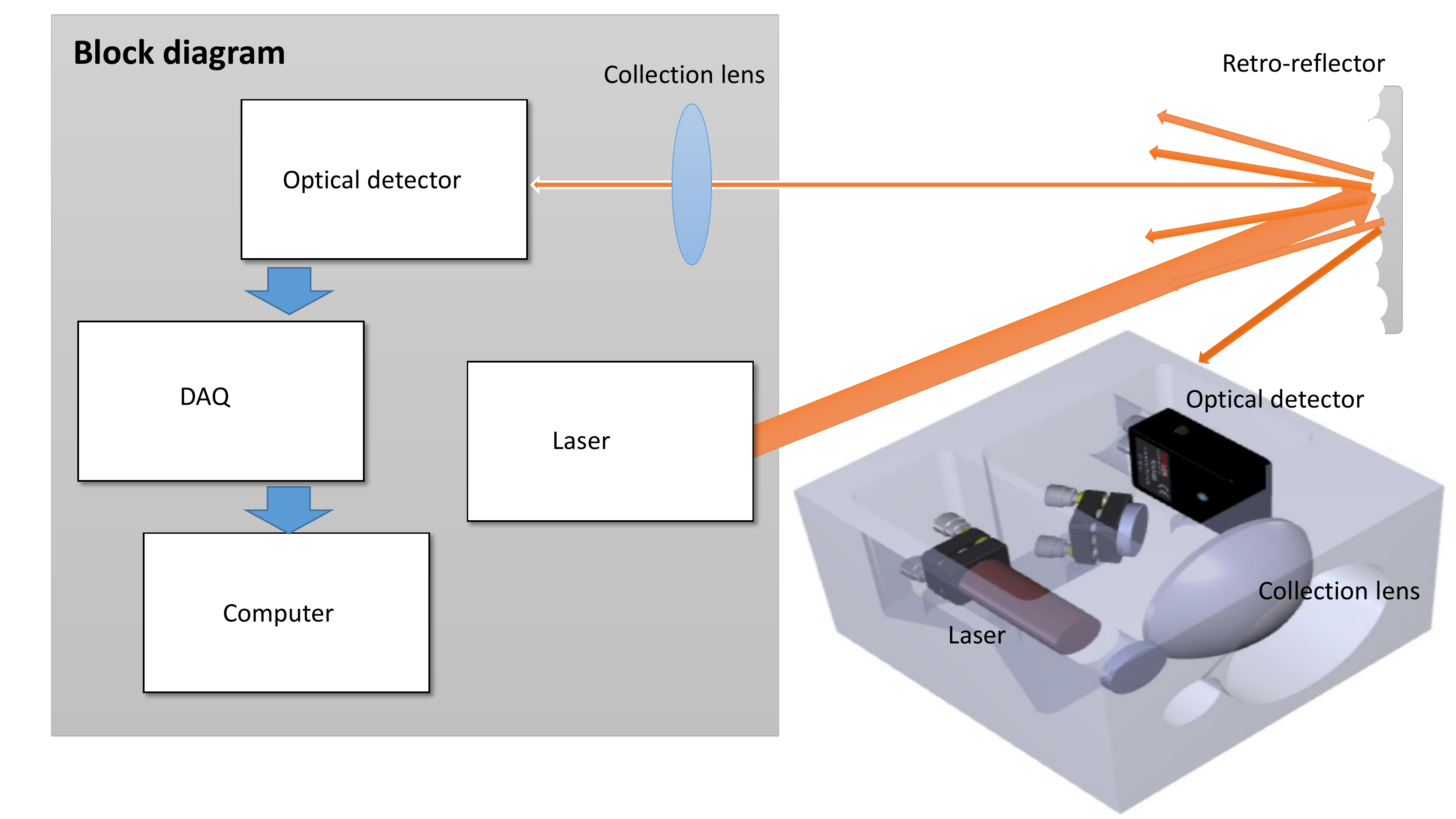}}
\caption{Block diagram of the sensor head and SOLIDWORKS drawings of the prototype sensor head. The standalone Linux-based Red Pitaya system with a Python application is not shown.}
\label{fig1}
\end{figure}

The experimental setup is shown in Figure \ref{fig1}. The optical hardware of the setup consists of a 8 mW 650 nm diode laser, a 50.8 mm collection lens with a focal length of 100 mm. The laser beam was collimated and transmitted through the monitoring region, where it hits a rough surface (reflective tape ifm-E21015) generating a speckle pattern. The retro-reflected speckle pattern is then focused by the 50.8 mm lens and detected  with a PIN photo-diode with a bandwidth of 15 MHz. The laser control and data acquisition is performed with a standalone Linux-based Red Pitaya system with a Python application. In the following we demonstrate the capability of the fire detector in a laboratory environment and at a factory site (Energnist I/S waste plant in Kolding, Denmark). 

\subsection{Laboratory environment }

\begin{figure}[htbp]
\centering
\fbox{\includegraphics[width=\linewidth]{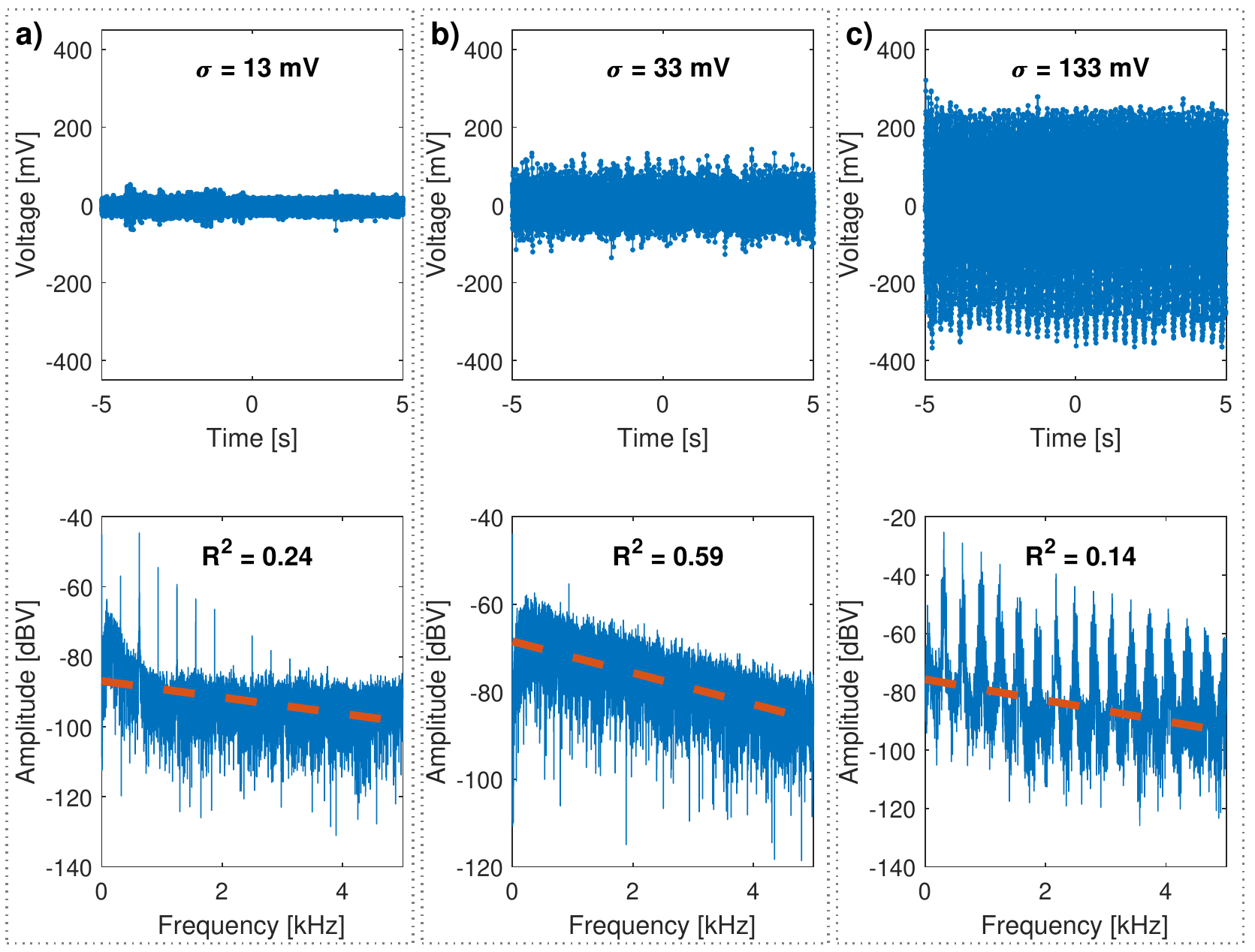}}
\caption{Recorded data in laboratory environment. Range is 101 meters from sensor to retroreflector. Heat source (a heat gun) placed halfway between laser and retroreflector, 30 cm under the beam. The figures shows voltage time traces for 10 seconds measurement time and the associated Fourier transformed noise spectrum for a) with no heat source ( heat gun), b) with heat gun and c) without heat gun, but with heavy mechanical vibration of the sensor. The red line is a linear fit to the Fourier transformed noise data. The time domain variance, $\sigma^2 $, and the  R$^2$ of the linear fit in frequency domain are two descriptors that can be indicative of a fire. Corresponding values are shown on the figure.}
\label{fig3}
\end{figure}

\begin{figure}[htbp]
\centering
\fbox{\includegraphics[width=\linewidth]{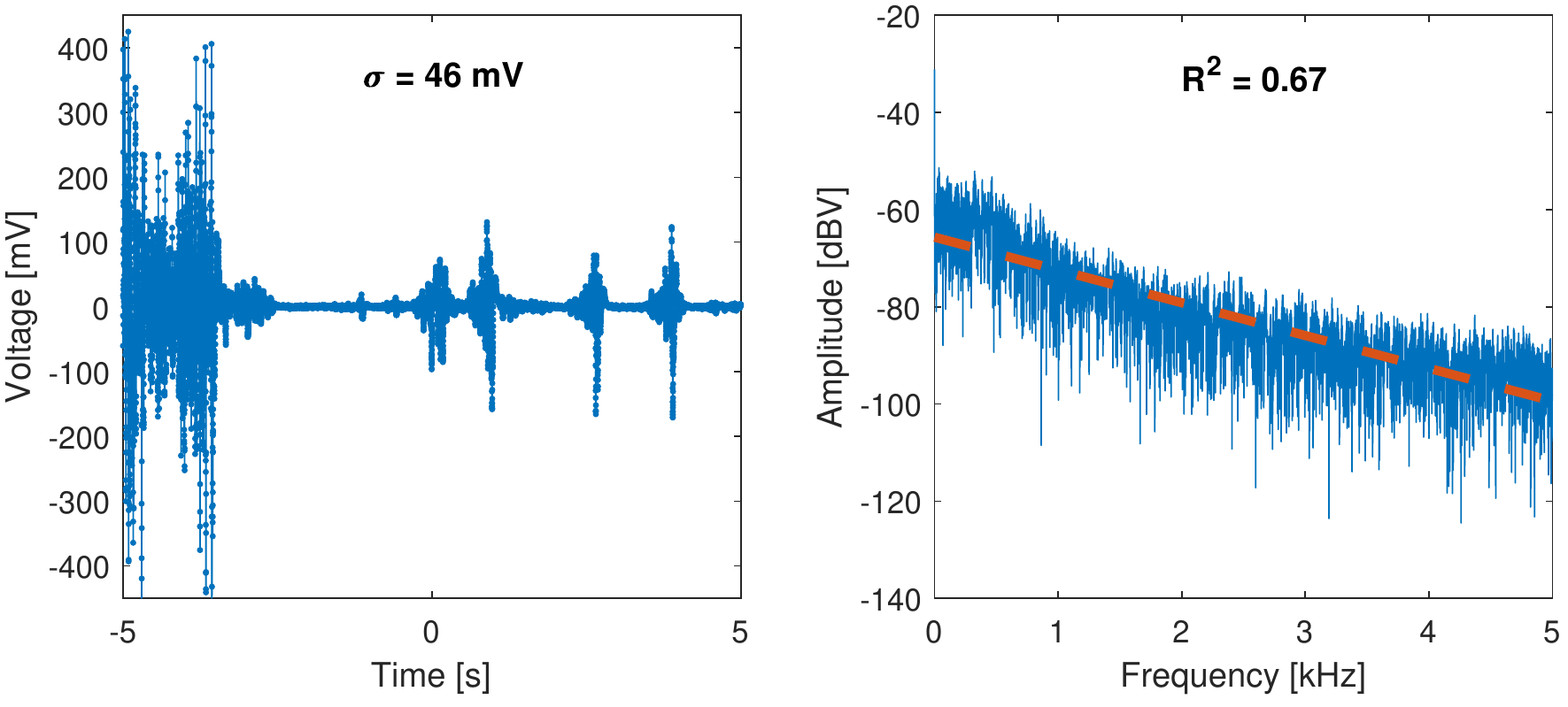}}
\caption{Recorded data in an outdoor environment  when smoke/dust is obscuring the light beam. Range is 10.5 meters from sensor to retroreflector. Heat source (a heat gun) placed halfway between laser and retroreflector, 30 cm under the beam. The figures shows voltage time traces for 10 seconds measurement time and the associated Fourier transformed noise spectrum. The red line is a linear fit to the Fourier transformed noise data. Values of the voltage standard deviation and $ R^2 $ of the linear fit of the spectrum is given.} 
\label{fig3a}
\end{figure}

In the following we present results in sensing with a range of 101 meters (limited by the size of the building). The first measurements are conducted in the controlled laboratory environment, where all parameters and noise sources are completely controlled. These measurements act as validation and calibration of the sensor system. In this case the sensor can simply recognize a fire by measuring the noise variance and perform a linear fit of the Fourier transformed (FFT) noise spectrum. The data is shown in Figure \ref{fig3} for three different conditions; no influence, heat gun and mechanical vibration. The recorded voltage time traces are shown in Figure \ref{fig3}) in a predefined time interval, e.g. 10 second time windows, and the calculated corresponding Fourier transformed noise spectrum are shown in Figure \ref{fig3}. The noise power in a broad frequency range is calculated and a linear fit is applied to the data. Broad frequency noise (from 10 Hz - 5 kHz) with a linear trend is an indication of fire, while on the other hand narrow linewidth noise sources are indication of mechanical resonances and needs to be filtered out. Figure \ref{fig3}a) shows the noise spectrum of the time trace without a heat gun. The red line is a linear fit to the FFT noise data with coefficient of determination R${^2}$ =0.24. The two parameters, the linear fit and the variance of the time trace, can be used as a first indication of fire. For instance the system could be set to give an alarm when the variance is greater than 400 $ \mathrm{mV}^2 $ and the linear fit is better than R${^2}$ > 0.45. Figure \ref{fig3}b) shows data where a heat gun is used to simulate a fire. The heat source (a heat gun) placed halfway between laser and retroreflector, 30 cm below the laser beam. The noise spectrum (blue FFT trace) has a much higher degree of linearity (R${^2}$ = 0.59) and the noise signal is approximately 25 dB above the case of no heat source. In this case the system triggers automatically an alarm. However, without a heat source we find, not surprisingly, that mechanical vibration gives a very noisy time trace, such that the variance is above the threshold for fire. However, the linear fitting function gives a relatively low R${^2}$ value of 0.14, thus the system can discard that the noise is due to a fire. Note that the applied mechanical vibration had a very high amplitude, probably much higher than in real life applications in a factory setting. In order to estimate the influence of pressure changes tests were also performed with the heat gun working as a blower. In this case no significant noise signal was measured, which was also expected from equation \ref{eq:2}.

One of the main features of our sensor is that it is measuring the noise spectrum and not the absolute back-reflected optical power. This makes our fire detection technology possible to detect fire in very dusty areas, which are prone to fires, where current systems show limited performance. This is tested by changing the transmission of the optical channel due to smoke and dust. Figure \ref{fig3a} shows an example of the results obtained when obstructing the optical channel with smoke. We clearly see that the time trace has parts where no light is transmitted through the channel, however since each measurement is taking 10 second we still gain some information about the convective heat changes. Both the linear fit and the variance gives the signature of a fire, so the system gives an alarm.

\subsection{Factory acceptance test}

\begin{figure}[htbp]
\centering
\fbox{\includegraphics[width=\linewidth]{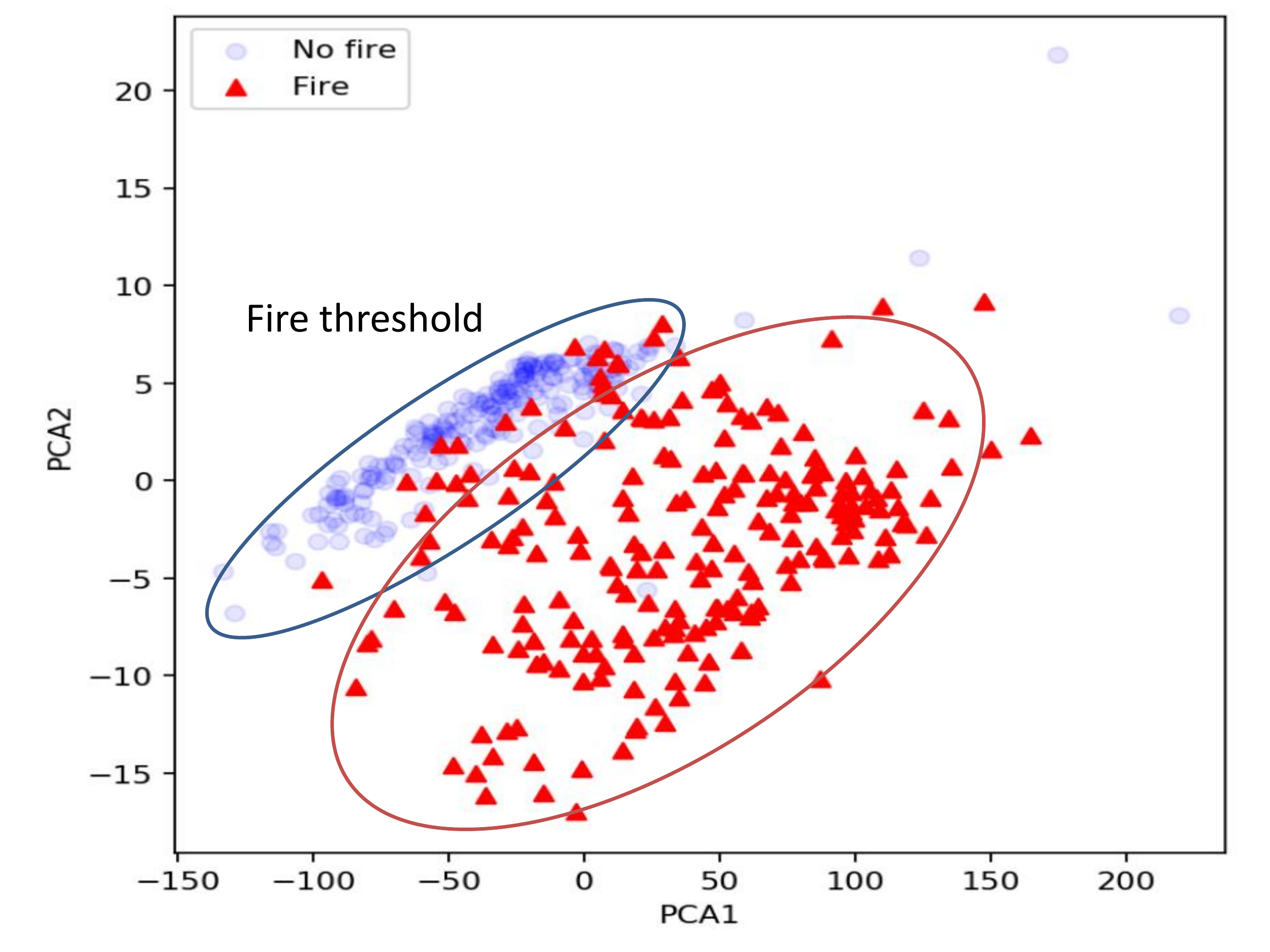}}
\caption{Test results of nonlinear support vector machine classifier when trained on only the two most significant principal components derived from a total of 8 descriptors. The principal component analysis (PCA) from a total of 146306 samples with 221 artificially simulated fires. The two principal components plotted are the two components with the highest variation in the data, i.e. the components are ordered by significance. The samples shown in the plot is a subset of the total acquired data. The complete dataset has been split into a training and a test dataset.}
\label{fig5}
\end{figure}

In order to validate the fire detector we have tested it at the Energnist I/S waste plant in Kolding, Denmark \cite{Energnist.dk}. The environment at the plant is extremely harsh, noisy and dusty and the plant normally has 3-4 false alarms per month. To conduct the test a prototype of the system based on the Red Pitaya development-kit with a 3G modem for remote administration was implemented. All packaged neatly in a generic surveillance camera case. A Python application running on Linux based Red Pitaya, sampled an analog input every 3 seconds (1600 raw samples). By the end of every sampling cycle a preprocessing step extracted 8 descriptors which were stored in a MySQL database also running on the device. These descriptors are the variance and Frobenius norm in both time and frequency domain in addition to the 99th percentile in frequency domain and the coefficients of a linear fit of the spectrum with a measure of goodness of fit. The descriptors form a 8-dimensional feature space, which can then be used for classification. As a means of filtering out unimportant features, principal component analysis (PCA) is used for feature selection. Based on those principal components, we use a support vector machine (SVM) with radial basis function kernel to obtain a binary classifier with nonlinear decision boundary \cite{hastie2005elements,gunn1998support}. A total of 146306 samples were acquired during the test, where 221 of those were artificially simulated fires. To get those samples a heat gun pointing diagonally from below the sensor in the direction of the beam was used. Figure \ref{fig5} shows results obtained when evaluating the classifier on a subset of the data reserved for testing. The data is transformed into the basis spanned by the two most significant principal components derived from all of the 8 descriptors. The accuracy of detecting a fire is 91$\%$, which is a very good result taking into account the harsh environment. However, we can improve the accuracy to 94$\%$ by including the third most significant PCA component. Although a high accuracy was achieved, it must be noted that fire samples were generated with a heat gun. A real live scenario with a developing fire will certainly not result in such separable data. Note that the accuracy of the SVM classifier can also be improved by using redundancy to lower the uncertainty from individual predictions, thus reducing the risk of false alarms. By performing N classifications and then using the most common class as the final predicted class (a majority vote) the misclassification rate can be lowered substantially as described by the cumulative distribution function of a binomial distribution. For example, the misclassification rate of 6$\%$ above could be reduced to 1$\%$ for N = 3, or 0.2$\%$ for N = 5. This makes the sensor more conservative in its prediction and increases the response time by a factor of N.

To get an unbiased estimate, a set of extra parameters should be introduced into future field tests: Heat/fire (source) intensity, preferably something simulating a developing fire. However in the case with waste plant, such a variation in fire simulation was not possible because of physical constraints as the laser beam was pointing across a large inaccessible area. The fire detection system achieved an uptime of 9 days, with the only problem encountered with the storage on the micro SD card, as they are not designed for frequent writes. If internet connectivity is desired for such a system, an embedded solution should have an on-board implementation that is accessible from the operating system itself. In our case, the 3G modem was attached to an external wifi-module making it impossible to achieve fine-grained control of the connection and in particular, reconnection when the signal is lost.

\section{Conclusion}

We have shown that the fire detector system with an integrated machine learning tool can detect small fires even in very noisy and dusty environment, such as the Energnist I/S waste plant in Kolding, Denmark. We believe that the direct impact of the technology could be seen in the costs associated with false alarms, as this will be drastically reduced for the benefit of companies and society. The fire detection system can improve the fire safety of old infrastructures and machinery without the need of further investments for total renovation. New application of the detection system may pave the way for early detection of fires in a number of environments where traditional fire detectors cannot be installed opening up a significant future market potential. We have demonstrated that the system could detect fires in the range of up to more than 100 meters.

In future work we will change the laser with a more powerful one in order to increase the sensing range. For example, using one with 25-30 mW of optical power and with a slightly bigger collection lens with a 75 mm diameter. This should make the finale prototype sensor more than 5 times as sensitive, thus we would be able of expanding the sensing range. So that the finale prototype system would be able to perform fire detection sensing in a length range of more than 500-600 meters with the same SNR as our current system. Further improvements could be done on the optical detector hereby achieving length of more than 1 km. While the laser, lens and sensor worked impeccably, the Red Pitaya board itself is not an optimal choice for a production system. First, it is designed as a testing/development/prototyping board and is missing features present in industrial-level embedded solutions: solid-state storage, advanced debugging capabilities, watchdog circuitry etc. The system had an uptime of 9 days and was only limited by the SD card. Problems with the storage on the micro SD card was encountered as they are not designed for frequent writes. Finally, if internet connectivity is desired for such a system, an embedded solution should have an on-board implementation that is accessible from the operating system itself. In our case, the 3G modem was attached to an external Wi-Fi module making it impossible to achieve fine-grained control of the connection and in particular, reconnection when the signal was lost. Better performance and higher sampling rates are to be achieved by rewriting parts of the application in a lower-level language (in C, C++ or Rust). In particular, linear regression used to generate a subset of the features used for prediction is an expensive operation when done in a dynamic language such as Python. By moving this and other feature extraction tasks (and even the sampling task itself) into a separate process, multitasking and multicore capabilities of the underlying system can be utilized. A move like this is not without a downside, as it may complicate the overall architecture of the application, but higher stability and lower CPU usage will outweigh the added complexity.

\section{Funding Information and acknowledgement}

The project was supported by funds from the Danish Agency for Institutions and Educational Grants and Eurostars project E9105 - Firedetect. We would like to thank Lap-Sikkerhed (Denmark) for many fruitful discussions on fire-safety and fire-detectors.

\end{document}